\documentclass[manuscript, nonacm]{acmart}

\usepackage{array}
\usepackage{pdflscape}
\usepackage{longtable}
\usepackage{rotating}
\usepackage{graphicx}
\usepackage{vcell}
\usepackage{siunitx}
\usepackage{tabu}
\usepackage{lscape}
\usepackage{afterpage}

\AtBeginDocument{%
  \providecommand\BibTeX{{%
    \normalfont B\kern-0.5em{\scshape i\kern-0.25em b}\kern-0.8em\TeX}}}

\begin{document}

\title{Emotion Recognition among Couples: A Survey}

\author{George Boateng}
\email{gboateng@ethz.ch}
\affiliation{%
  \institution{ETH Z{\"u}rich}
  \city{Zurich}
  \country{Switzerland}
}

\author{Elgar Fleisch}
\email{efleisch@ethz.ch}
\affiliation{%
  \institution{ETH Z{\"u}rich}
  \city{Zurich}
  \country{Switzerland}
}
\affiliation{%
 \institution{University of St. Gallen}
 \city{St. Gallen}
 \country{Switzerland}
}

\author{Tobias Kowatsch}
\email{tkowatsch@ethz.ch}
\affiliation{%
  \institution{ETH Z{\"u}rich}
  \city{Zurich}
  \country{Switzerland}
}

\affiliation{%
 \institution{University of St. Gallen}
 \city{St. Gallen}
 \country{Switzerland}
}

\renewcommand{\shortauthors}{Boateng et al.}

\begin{abstract}
Couples’ relationships affect the physical health and emotional well-being of partners. Automatically recognizing each partner’s emotions could give a better understanding of their individual emotional well-being, enable interventions and provide clinical benefits. In the paper, we summarize and synthesize works that have focused on developing and evaluating systems to automatically recognize the emotions of each partner based on couples’ interaction or conversation contexts. We identified 28 articles from IEEE, ACM, Web of Science, and Google Scholar that were published between 2010 and 2021. We detail the datasets, features, algorithms, evaluation, and results of each work as well as present main themes. We also discuss current challenges, research gaps and propose future research directions. In summary, most works have used audio data collected from the lab with annotations done by external experts and used supervised machine learning approaches for binary classification of positive and negative affect. Performance results leave room for improvement with significant research gaps such as no recognition using data from daily life. This survey will enable new researchers to get an overview of this field and eventually enable the development of emotion recognition systems to inform interventions to improve the emotional well-being of couples.
\end{abstract}



\maketitle

\section{Introduction}
The emotions experienced by romantic partners are linked with relationship quality and the management of chronic diseases. Couples’ emotions experienced during conflicts predict if these couples stay together in the long term (\cite{gottman2014}). For example, couples heading for break-up show more negative emotions and less positive emotions than happy couples \cite{carstensen1995, gottman2005}. For couples with one partner having a chronic disease, the burden of the disease management is shared by both partners and it takes a toll on the emotional well-being of not just the patient but also on the supporting partner. Furthermore, social support from partners in chronic disease management has been shown to either have positive or negative effects on the emotional well-being of patients \cite{prati2010, iida2010, bolger2007}. Because of the importance of emotions among couples, researchers are working towards understanding the emotional processes that take place in intimate relationships (e.g., \cite{farrell2018, smith2019}) and the link between emotions and social support in couples’ dyadic management of chronic diseases \cite{luescher2019}. Consequently, being able to automatically recognize each partner’s emotions could enable the research of social and health psychologists, and also inform the development of dyadic interventions (where partners are both involved e.g., \cite{ketcher2021}) to improve the emotional well-being, relationship quality, and chronic disease management of couples.

Emotion recognition among couples entails the recognition of the emotions of each romantic partner based on the context of their interaction. This task has a number of differences and similarities with other kinds of emotion recognition tasks. Standard emotion recognition attempts to recognize the emotion of each individual and uses various kinds of stimuli to induce an emotional reaction such as driving \cite{zepf2020}, listening to music or watching a movie \cite{abadi2015}, giving a speech \cite{schmidt2018} and engaging in a conversation with another person  \cite{poria2019}. 

The stimuli used for couples’ emotion recognition is a conversational context and hence, it has some similarities to emotion recognition tasks based on individuals having a conversation. The conversational context has the unique challenge of turn-taking dynamics which requires the system to correctly identify who is speaking for each speech segment, termed speaker diarization, consequently making this kind of emotion recognition task more challenging than others that employ stimuli such as listening to music or watching a video.

Most emotion recognition tasks using conversational contexts tend to use actors acting out hypothetical scenarios \cite{busso2008, metallinou2010, busso2016}. Actors tend to exaggerate their emotional expressions and the models trained with that type of data have been shown to perform better than models trained on real-world data  \cite{dmello2015}. It is not clear if such emotion recognition systems will perform well on data from non-actors. On the other hand, some emotion recognition tasks employ people (e.g., 2 strangers) having real conversations \cite{grimm2008, park2020}. This type of emotion recognition is the closest to couples’ emotion recognition in terms of the conversational context and the dyadic and realistic nature of the interaction. However, for the couples’ context, because the two individuals involved are in a romantic relationship, various insights from psychology research about couples can be leveraged to better recognize the emotion of each partner. For example, partners tend to influence each other’s emotions throughout an interaction \cite{berscheid2001}, and hence various interpersonal dynamics could be leveraged to adequately recognize each partner’s emotions (e.g.,  \cite{chakravarthula2018, boateng2021}). 

Because of the uniqueness of couples’ emotion recognition and the potential clinical utility, there is a need to synthesize emotion recognition approaches focused on the couples’ context. Several works have developed systems to automatically recognize the emotions of couples. In this paper, we describe and discuss these works and give a comprehensive overview of this research field. We surveyed 28 articles published over the past 11 years (2010 - 2021) given the first set of works on this topic were published in 2010. We detail the datasets, features, algorithms, evaluation, and results of each work as well as present main themes. We also discuss current challenges, research gaps and propose future research directions. There are surveys of emotion recognition works focused on specific modalities such as visual and speech modalities (\cite{zeng2008}), wearables (\cite{schmidt2019}), multimodality (\cite{dmello2015, poria2017}) and contexts such as driving (\cite{zepf2020}) and conversation (\cite{poria2019}). However, this is the first survey of works that focus on emotion recognition within the context of couples’ interactions or conversations.

The rest of this paper is organized as follows. In Section 2, we describe the scope of the survey and the approach used to select the papers. In Section 3, we give an overview of the surveyed works. In Section 4, we describe emotion models, elicitation, and annotation approaches that have been used. In Section 5, we give an overview of the datasets that have been used in the surveyed works. In Section 6, we describe the modalities used, how they have been preprocessed, and the features extracted from them. In Section 7, we discuss the algorithms that have been used, modeling approaches considering the unique context of couples’ interactions, evaluation approaches, and the results obtained. In Section 8, we discuss research gaps, challenges, and future directions. We conclude in Section 9.

\section{Survey Scope and Methodology}
Our methodology is similar to those of Zepf et al. \cite{zepf2020}. We sought to survey papers that used data to automatically recognize the emotions of each romantic partner based on the couple’s interaction or conversation context. We developed a list of search terms that covered three concepts: (1) emotions/emotional behavior (emotion, affect, affective, moods, behavior), (2) recognition (recognition, prediction, classification, behavior signal processing, affective computing, machine learning, deep learning, neural networks) and (3) couples (couples, dyad, spouse, married). We entered the search terms into the following databases: IEEE, ACM, Web of Science, and Google Scholar. We also looked through the references of relevant papers. 

To be included, the papers had to perform automatic recognition of each partner’s emotions (e.g., positive or negative valence), emotional behavior or emotional states (e.g., positive affect/positivity, negative affect/negativity, sadness), employed statistical or machine learning approaches, and used data collected from the context of couples’ interaction or conversations. We included papers that were not peer-reviewed yet but were on archival databases such as arXiv for completeness. 

We excluded papers that used data from interacting partners that are not real couples, i.e., individuals acting out dyadic interactions either using a script or engaging in spontaneous sessions such as the following datasets \cite{muller2015,  busso2008, metallinou2010, busso2016}. We also excluded papers that recognized couple behavior which are not emotional states such as level of blame \cite{black2011}, conflict \cite{timmons2017}, suicidal risk \cite{chakravarthula2020}. In particular, we excluded one paper \cite{bari2020} that focused on recognizing stressful conversations from other kinds of stressful situations among couples. We also excluded papers that recognized couples’ relationship state (e.g., happy vs sad couple) \cite{uhlich2021} rather than each partner’s emotions. Another related paper that we excluded is \cite{boateng2020d} which proposed a research plan for emotion recognition but does not present performed analysis and results in the paper. After using these inclusion and exclusion criteria, we had 28 relevant articles published between 2010 and 2021 (Table \ref{tab:studies_overview}).

\section{Overview of Works}
Out of the 28 surveyed papers, a majority of the works (n=24) have been done primarily by the Signal Analysis and Interpretation Laboratory (SAIL) team at the University of Southern California which published the first set of works on this topic in 2010 \cite{black2010, lee2010}. Subsequent works extended or built upon previous works from the research group. Few works have been done by researchers outside this research lab and include the following \cite{crangle2019, boateng2020a, boateng2021, biggiogera2021}.

Together, these works have used 5 datasets that were collected from a laboratory setting (Table \ref{tab:studies_datasets}). Most of these works used emotion labels from external raters with few using self-reported data from the couples themselves. Only three modalities have been used — acoustic, lexical, and visual — with acoustic being the most used modality. Multimodal fusion has been done mostly for acoustic and lexical modalities using feature-level and decision-level fusion. Support vector machines are the most used algorithm. Various intrapersonal considerations (e.g., saliency) and interpersonal considerations (e.g., synchrony) have been leveraged (details in a future section). Evaluations have mostly been done with leave-one-couple-out cross-validation with accuracy being used as the metric.

\begin{landscape}
\scriptsize
\centering
\begin{longtable}{|>{\hspace{0pt}}m{0.07\linewidth}|>{\hspace{0pt}}m{0.03\linewidth}|>{\hspace{0pt}}m{0.081\linewidth}|>{\hspace{0pt}}m{0.07\linewidth}|>{\hspace{0pt}}m{0.03\linewidth}|>{\hspace{0pt}}m{0.1\linewidth}|>{\hspace{0pt}}m{0.03\linewidth}|>{\hspace{0pt}}m{0.06\linewidth}|>{\hspace{0pt}}m{0.06\linewidth}|>{\hspace{0pt}}m{0.06\linewidth}|>{\hspace{0pt}}m{0.127\linewidth}|} 
\caption{Overview of datasets\label{tab:studies_datasets}}\\ 
\hline
\textbf{Dataset} & \textbf{No of Couples} & \textbf{Couples Context} & \textbf{Data} & \textbf{Hours of Data} & \textbf{Session Type} & \textbf{Mins per Session} & \textbf{Annotation Type} & \textbf{Annotation Scope} & \textbf{Emotion Model} & \textbf{Emotions} \endhead
\hline
UCLA / UW Couples Therapy & 134 & English-speaking, Chronically distressed couples & Video, audio, transcripts  & 96 & Relationship problem (1) & 10 & Observer & Global & Dimensional, Categorical & Positive affect, Negative affect, Sadness, Anger, Anxiety \\ 
\hline
Moffit Center Cancer Conversation & 85 & English-speaking, Couples managing cancer & Video, audio, transcripts & 27 & Neutral (1), Cancer management issue (1) & 10 & Observer & Local (utterance / speaker-turn) & Categorical & Hostile (Negative), Positive, Constructive (Neutral) \\ 
\hline
KU Leuven Dyadic Interaction & 101 & Dutch-speaking Couples & Video, audio & 51 & Neutral (1), Positive (1), Negative (1) & 10 & Self & Global, Local (continuous) & Categorical, Dimensional & Categorical: anger, sadness, anxiety, relaxation, happiness, Dimensional: Valence and Arousal \\ 
\hline
Stanford Psychotherapy & 3 & English-speaking Couples Therapy & Video, audio, transcripts & 18 & Relationship discussion (1) & 60 & Observer & Local (utterance) & Categorical & Anger, sadness, joy, tension, neutral \\ 
\hline
UZH Couples Interaction & 368 & German-speaking couples in Switzerland & Video, audio, transcripts & 637 & Conflict discussion (1), Mutual support discussion (2) & 8 & Self, Observer & Global, Local (utterance) & Dimensional & Positive, negative, happy vs sad, good mood vs bad mood, relaxed vs angry, calm vs stressed \\
\hline
\end{longtable}
\end{landscape}


\begin{landscape}
\scriptsize
\centering
\begin{longtable}{|>{\hspace{0pt}}m{0.07\linewidth}|>{\hspace{0pt}}m{0.07\linewidth}|>{\hspace{0pt}}m{0.10\linewidth}|>{\hspace{0pt}}m{0.12\linewidth}|>{\hspace{0pt}}m{0.07\linewidth}|>{\hspace{0pt}}m{0.06\linewidth}|>{\hspace{0pt}}m{0.08\linewidth}|>{\hspace{0pt}}m{0.04\linewidth}|>{\hspace{0pt}}m{0.03\linewidth}|>{\hspace{0pt}}m{0.02\linewidth}|>{\hspace{0pt}}m{0.13\linewidth}|}
\caption{Studies focused on emotion recognition among couples. ACC=Accuracy, Corr=Spearman Correlation, CV=cross validation, DD=Diversity Density, DNN=Deep neural network, GMM=Gaussian Mixture Models, HMM=Hidden Markov Model, LDA=Linear discriminant analysis,  LR=Logistic Regression,  LNCO=Leave-n-couple-out, LOCO= Leave-one-couple-out,  MAE=Mean Average Error, ML=Maximum Likelihood, MM=Markov Model, RBF=Radial Basis Function, RF=Random Forest, SPRT=Sequential Probability Ratio Test, SVM=Support vector machine, UAR=Unweighted Average Recall}
\label{tab:studies_overview}\\
\hline
\textbf{Ref} & \textbf{Dataset} & \textbf{Modalities} & \textbf{Features} & \textbf{Interpersonal} & \textbf{Intrapersonal} & \textbf{Algorithms} & \textbf{Evaluation} & \textbf{Metric} & \textbf{Class} & \textbf{Main Best Results} \\ \hline
\endhead
Biggiogera et al., 2021 \cite{biggiogera2021} & UZH Couples Interaction & Acoustic (A), Lexical (L) & A: Prosodic and spectral eGeMAPS features, L: Ngram + TFIDF, LIWC, Deep sentence embeddings & No & No & Linear SVM & 10-fold CV couple disjoint & UAR & 2 & Negative vs Positive affect: 69.2\% \\ 
\hline
Black et al., 2010 \cite{black2010} & UCLA / UW Couples Therapy & Acoustic & Prosodic and spectral & No & No & Linear SVM, LDA & LOCO CV & ACC & 2 & Positive affect: 82\% (female), 75\% (male), Negative affect: 77\% (female), 76\% (male) \\ 
\hline
Black et al., 2013 \cite{black2013} & UCLA / UW Couples Therapy & Acoustic & Prosodic and spectral & No & No & Linear SVM, LR & LOCO CV & ACC & 2 & Positive affect: 77.9\% (female), 72.9 (male), Negative affect: 80\% (female), 85.7\% (male) \\ 
\hline
Boateng et al., 2020 \cite{boateng2020a} & KU Leuven Dyadic Interaction & Acoustic & Deep acoustic embeddings (Spectrograms + Pretrained CNN) & No & Yes & Linear SVM & LOCO CV & UAR & 2 & Negative vs Positive valence: 74.8\% (female), 53.3\% (male) \\ 
\hline
Boateng et al., 2021 \cite{boateng2021} & UZH Couples Interaction & Acoustic (A), Lexical (L); Fusion: Feature level & A: Prosodic and spectral eGeMAPS features, L: Deep sentence embeddings & Yes (Dyadic Influence) & No & Linear SVM, RBF SVM, RF & 10-fold CV couple disjoint & UAR & 2 & Negative vs Positive valence: 64.8\% (female), 56.1\% (male) \\ 
\hline
Chakravarthula et al., 2015 \cite{chakravarthula2015} & UCLA / UW Couples Therapy & Lexical & Unigram & No & Yes (Dynamic modeling) & ML as baseline, HMM & LOCO CV & ACC & 2 & Negative affect: 88.57\% (female), 83.57\% (male) \\ 
\hline
Chakravarthula et al., 2018 \cite{chakravarthula2018} & UCLA / UW Couples Therapy & Lexical & Ngram & Yes (Dyadic Influence) & Yes (Dynamic modeling) & ML as baseline, HMM & LOCO CV & ACC & 2 & Negative affect: 88.93\% \\ 
\hline
Chakravarthula et al., 2019 \cite{chakravarthula2019} & Moffit Center Cancer Conversation & Acoustic, Lexical; Fusion: Feature level, Decision level & Acoustic: Prosodic and spectral eGeMAPS featuresLexical: Deep senstence embedding & No & No & DNN & LOCO CV & UAR & 3 & Positive vs Negative vs Neutral: 57.42\% \\ 
\hline
Chakravarthula et al., 2021 \cite{chakravarthula2021} & UCLA / UW Couples Therapy & Lexical & N-gram, ELMo & No & No & ML, GRU & 6-fold CV couple disjoint & Corr & N/A & Positive affect: 0.5, Negative affect: 0.58, Anxiety: 0.18,Anger: 0.52, Sadness: 0.34 \\ 
\hline
Crangle et al., 2019 \cite{crangle2019} & Stanford Psychotherapy & Acoustic & Prosodic and spectral features & No & No & RF & Hold out & ACC & 5 & Anger, sadness, joy, tension, neutral, Couple A: 87\% (male), 90\% (female), Couple B: 78\% (male), 84\% (female), Couple C: 95\% (male), 88\% (female) \\ 
\hline
Georgiou et al., 2011 \cite{georgiou2011} & UCLA / UW Couples Therapy & Lexical & Unigram & No & No & ML & LOCO CV & ACC & 2 & Positive affect: 88.9\%, Negative affect: 86.7\%, Sadness: 61.6\% \\ 
\hline
Gibson et al., 2011 \cite{gibson2011} & UCLA / UW Couples Therapy & Acoustic & Spectral & No & Yes (Sailency) & RBF SVM as baseline, DD SVM & 10-fold CV couple disjoint & ACC & 2 & Positive affect: 74.3\% (female), 58.6 (male), Negative affect: 77.9\% (female), 71.4\% (male), Sadness: 66.4\% (female), 63.6\% (male) \\ 
\hline
Gibson et al., 2015 \cite{gibson2015} & UCLA / UW Couples Therapy & Acoustic (A), Lexical (L), Visual (V); Fusion: Feature level, Decision level & A: Prosodic and spectral, L: TFIDF, V: Power spectral density (PSD) of head motion vectors & No & Yes (Sailency) & DD, Linear SVM & LOCO CV & ACC & 2 & Positive affect: 65.13\% (all modalities), Negative affect: 70.34\% (lexical) \\ 
\hline
Katsamanis et al., 2011 \cite{katsamanis2011} & UCLA / UW Couples Therapy & Acoustic, Lexical; Fusion: Unclear & MFCC, TFIDF & No & Yes (Sailency) & RBF SVM as baseline, DD SVM. & 10-fold CV couple disjoint & ACC & 2 & Positive affect: 93\% (lexical), Negative affect: 95\% (lexical), Sadness: 80\% (lexical) \\ 
\hline
Lee et al., 2010 \cite{lee2010} & UCLA / UW Couples Therapy & Acoustic & Prosodic & Yes (Synchrony) & No & MM & LOCO CV & ACC & 2 & Positive vs negative affect: 76\% \\ 
\hline
Lee et al., 2011a \cite{lee2011a} & UCLA / UW Couples Therapy & Acoustic & Prosodic and spectral & Yes (Synchrony) & No & RBF SVM & LOCO CV & ACC & 2 & Positive vs negative affect: 51.79\% \\ 
\hline
Lee et al., 2011b \cite{lee2011b} & UCLA / UW Couples Therapy & Acoustic & Prosodic and spectral & Yes (Synchrony) & Yes (Sailency) & DD & LOCO CV & ACC & 2 & Positive vs negative affect: 53.93\% \\ 
\hline
Lee et al., 2012 \cite{lee2012} & UCLA / UW Couples Therapy & Lexical & TFIDF & No & Yes (Sailency) & DD, SPRT & 10-fold CV couple disjoint & ACC & 2 & Positive affect: 76.1\%, Negative affect: 74.2\%, Sadness: 54.2\% \\ 
\hline
Lee et al., 2014 \cite{lee2014} & UCLA / UW Couples Therapy & Acoustic & Prosodic and spectral & Yes (Synchrony) & No & HMM, Factorial HMM & LOCO CV & ACC & 2 & Positive vs negative affect: 62.86\% \\ 
\hline
Li et al., 2016 \cite{li2016} & UCLA / UW Couples Therapy & Acoustic & Prosodic and spectral features & No & No & SVM as baseline, DNN & LOCO CV & ACC & 2 & Negative affect: 77.14\% \\ 
\hline
Li et al., 2017 \cite{li2017} & UCLA / UW Couples Therapy & Acoustic & Prosodic and spectral & No & No & DNN (autoencoder) & LOCO CV & ACC & 2 & Negative affect: 69.64\%, Positive affect: 66.43 \% \\ 
\hline
Li et al., 2020 \cite{li2020} & UCLA / UW Couples Therapy & Acoustic & Prosodic and spectral, Deep acoustic embeddings & No & No & CNN, GRU & LNCO (n=4) & ACC & 2 & Positive affect: 65.36\%, Negative affect: 76.07\%, Sadness: 59.29\% \\ 
\hline
Tseng et al., 2016 \cite{tseng2016} & UCLA / UW Couples Therapy & Lexical & word2vec & No & No & ML as baseline, LSTM + RBF SVR & LOCO CV & ACC & 2 & Negative affect: 88.93\% \\ 
\hline
Tseng et al., 2017 \cite{tseng2017} & UCLA / UW Couples Therapy & Lexical & word2vec, Deep sentence embeddings & No & No & LSTM + RBF SVR & LOCO CV & MAE & N/A & Negative affect: 1.37 \\ 
\hline
Tseng et al., 2018 \cite{tseng2018} & UCLA / UW Couples Therapy & Acoustic (A), Lexical (L); Fusion: Feature level, Decision level, Gender based, Therapy stage & A: Prosodic and spectral features, L: Deep sentence embeddings & No & No & LSTM, DNN & LOCO CV & MAE & 2 & Negative affect: 1.22 \\ 
\hline
Tseng et al., 2019 \cite{tseng2019} & UCLA / UW Couples Therapy & Lexical & Deep sentence embeddings & No & No & LSTM & LOCO CV & ACC & 2 & Positive affect: 86.8\%Negative affect: 87.9\% \\ 
\hline
Xia el al., 2015 \cite{xia2015} & UCLA / UW Couples Therapy & Acoustic & Prosodic and spectral & No & Yes (Dynamic modeling) & SVM, LDA, Voted Perceptron as baseline, HMM + SVM, LDA, Voted Perceptron & LOCO CV & ACC & 2 & Positive affect: 81\% (female), 78\% (male), Negative affect: 79\% (female), 84\% (male), Sadness: 65\% (female), 61\% (male) \\ 
\hline
Xiao et al., 2015 \cite{xiao2015} & UCLA / UW Couples Therapy & Visual & Line Spectral Frequencies of head motion vectors & Yes (Synchrony) & No & GMM + Linear SVM & LOCO CV & ACC & 2 & Positive affect: 63\%, Negative affect: 57\% \\
\hline
\end{longtable}
\end{landscape}

\section{Background} 
In this section, we describe various emotion models and approaches to eliciting and annotating emotion data from couples.

\subsection{Emotion Models}
There are mainly two models of emotions used in the literature in emotion recognition: categorical and dimensional. Categorical emotions are based on the six basic emotions proposed by Ekman: happiness, sadness, fear, anger, disgust, and surprise  \cite{ekman1997}. Over time, additional emotion categories have been included and used in literature such as anxiety, frustration, etc. Dimensional approaches mainly use two dimensions: valence (pleasure) and arousal (activation) which are based on Russell’s circumplex model of emotions \cite{russell1980}. Valence refers to how negative to positive the person feels and arousal refers to how sleepy to active a person feels. Using these two dimensions, several categorical emotions can be placed and grouped into the four quadrants: high arousal and negative valence (e.g., stressed, angry), low arousal and negative valence (e.g., depressed), low arousal and positive valence (e.g., relaxed), and high arousal and positive valence (e.g., excited).

\subsection{Elicitation}
Approaches for eliciting emotions in couples have generally happened in the lab/controlled settings and in daily life. In the lab, couples are asked to have emotionally charged conversations that are videotaped \cite{roberts2007}. Some of these conversations center on topics that cause distress in their relationship. These conversations elicit emotions during and after the conversations which are then annotated. In daily life, sensor data (e.g., audio) is collected from couples periodically \cite{robbins2014} or when conversation moments are detected \cite{boateng2020d}.

\subsection{Annotation}
Two approaches are used for emotion annotations by social psychologists: self-report and observer reports, and two scales: global and local (continuous, utterance-level). For self-reports, each partner provides emotion ratings right after the whole interaction/conversation (global/session ratings) with validated instruments such as the Affect Grid questionnaire \cite{russell1989} and Multidimensional Mood questionnaire \cite{steyer1997} or they are asked to watch a video recording of the conversation while providing continuous (moment-by-moment) emotion ratings using a joystick (e.g., \cite{roberts2007}). In the case of daily life, couples are periodically asked to complete self-reports such as the PANAS \cite{watson1988}, Affect Grid questionnaire \cite{russell1989} and Affective Slider \cite{betella2016} at random time \cite{sels2019b, sels2020, boateng2020d} or after sensor data recording \cite{boateng2020d}. Additionally, the dyadic nature of couples’ interactions enables the collection of partner-perceived emotions where each partner (e.g., partner A) is asked to provide a rating of their perception of the emotion of their partner (e.g., partner B) emotion after an interaction in the lab (e.g., sels2019a, sels2019b) or in daily life (e.g., sels2020). All these types of self-reports enable the collection of the subjective and perceived emotions of each partner. However, these ratings could be biased and may not reflect each partner’s actual emotions about the interaction. 

For observer reports, people are trained to watch the video recordings (e.g., in the case of lab data) and use a coding scheme to rate the interaction on specific emotional behaviors (e.g., SPAFF \cite{coan2007}) using continuous or utterance-level ratings (e.g., every 10 seconds or speaker turn) or global ratings (of the whole interaction). Such coding is also done for example, for audio data collected from couples’ daily life interactions \cite{robbins2014}. This manual coding process is costly and time-consuming as multiple coders need to be trained for this task \cite{kerig2004} and suffers from inter-rater reliability issues \cite{heyman2001, metallinou2013}. Furthermore, these ratings reflect the observers’ perceived emotions of the partners and they do not represent the subjective emotions of the partners.

\section{Studies and Datasets}
In this section, we describe all the datasets that have been used in the surveyed emotion recognition works, how they were collected and annotated. The surveyed papers used five (5) datasets, all of which were collected in the lab. Three (3) were observer annotated, one (1) was self annotated, and one (1) had both self and observer annotations (Table \ref{tab:studies_datasets}). The distribution of papers that have used the five datasets is as follows: UCLA/UW Couples Therapy (23), Cancer Conversation (1), KU Leuven Dyadic Interaction (1), Stanford Psychotherapy (1), and UZH Couples’ Interactions (2).

\subsection{UCLA/UW Couples Therapy}
Researchers conducted a longitudinal lab study at the University of California, Los Angeles, and the University of Washington in the U.S. with 134 seriously and chronically distressed heterosexual married couples \cite{black2010}. Their age statistics were as follows: the range is 22 to 72 years, the median age for men was 43 years (SD = 8.8), and the median age for women was 42 years (SD = 8.7). They were, on average, college-educated (median level of education for both men and women was 17 years, SD = 3.2). The sample was largely Caucasian (77\%), with 8\% African American, 5\% Asian or Pacific Islander, 5\% Latino/Latina, 1\% Native American, and 4\% Other. Couples were married for an average of 10.0 years (SD = 7.7) \cite{christensen2004}.

Couples received couples therapy for 1 year. They had conversations and discussed a problem in their relationship with no therapist or research staff present. They discussed the wife’s chosen topic for 10 minutes and the husband’s chosen topic for 10 minutes which were considered separate sessions. The sessions were recorded at three points in time: before the therapy, 26 weeks into it, and two years after the therapy sessions ended. There were 96 hours of data across 574 sessions. The sessions were videotaped and later transcribed and annotated by 3-4 trained coders. The annotators assigned 33 session-level (global) behavioral codes for each spouse on a scale of 1 - 9 using two coding schemes. The coding schemes are the Social Support Interaction Rating System (SSIRS) which consists of 20 codes that measure the emotional component of the interaction and the topic of conversation \cite{jones1998} and the Couples Interaction Rating System 2 (CIRS2)  which consist of 13 codes and were specifically designed for conversations involving a problem in a relationship \cite{heavey2002}. There were no local (utterance- or speaker-turn-level) annotations. 

Authors that used this dataset for emotion recognition tasks used only six codes in experiments due to low inter-evaluator agreement for the other codes. The codes that were used are level of blame, level of acceptance towards the other spouse, global positive affect, global negative affect, level of sadness, use of humor. They used the manual transcript of the data to automatically create word and speaker turn alignments which resulted in a smaller number of sessions and unique couples data used for the recognition experiments: 293 sessions. Also, they computed the mean values across raters and then selected data whose ratings were in the top 20\% and bottom 20\% for each of the codes. Consequently, the data used for analysis was from  60 - 85 unique husband/wife pairs. The task was cast as a binary classification for the two extremes for each code. In this survey, we consider works that used the affect-related codes: global positive affect, global negative affect, level of sadness, anxiety, and anger. 

\subsection{Moffit Center Cancer Conversation} 
Researchers conducted a study in which they collected data from 85 couples in the U.S. who were coping with advanced cancer (one spouse having cancer and the other being a caregiver) \cite{chakravarthula2019, reblin2018, reblin2019}. Here is the demographic data of the 82 couples whose data was eventually used: 29.3\% of patients were female and 70.1\% of caregivers were female, the mean age for patients was 66.8 years (SD = 9.2), and the mean age for caregivers was 64.8 years (SD = 9.4). On average, they had college or vocational education. The patient sample was largely Caucasian (92.7\%), with 6.1\% African American, 3.7\% Latino/Latina, 1.2\% Native American, and 0\% Other. The caregiver sample was also largely Caucasian (90.2\%), with 4.9\% African American, 3.7\% Latino/Latina, 2.4\% Native American, and 1.2\% Other. Couples were married for an average of 35 years (SD = 15.8) \cite{reblin2018}.

They engaged in a 10-minute neutral discussion (daily routine) and a 10-minute stressor discussion about an issue related to cancer management in controlled settings (e.g., clinic consult rooms, participant homes) with an experimenter present without facilitating. The issue was decided based on their ratings on Cancer Inventory of Problem Situations, \cite{heinrich1984} in which a list of 20 common cancer concerns (e.g., lack of energy, finances, over-protection) are rated as being not a problem, somewhat of a problem, or a severe problem. The interactions were audio-recorded. We estimated that a total of 27 hours of data was collected.

The audio was annotated on an utterance / speaker-turn level by 2 trained coders using the Rapid Marital Interaction Coding System, 2nd Edition \cite{heyman1995, heyman2004} with inter-rater reliability scores of Kappas above 0.88 for 20\% of all codes). Each utterance was assigned one behavioral code out of 7 codes which were then grouped into three: positive (low and high positive), hostile/negative (low and high hostile), neutral/constructive (constructive problem discussion). Additional codes were dysphoric affect and other which were not used for the recognition task. Hence, the task was framed as a 3-class classification problem. They used the manual transcripts to automatically create word, speaker turn and label alignments. This dataset has been used by \cite{chakravarthula2019}.

\subsection{KU Leuven Dyadic Interaction} 
Researchers conducted a Dyadic Interaction lab study in Leuven, Belgium with 101 heterosexual, Dutch-speaking couples \cite{boateng2020a, sels2019a, sels2019b}. The majority (n=96) cohabited and 7 were married. The average age was 26 years (SD=5), ranging from 18 to 53 years. The partners were together for 4.5 years (SD=2.8), ranging from 7 months to 21 years. There was no information about the ethnicity and education levels of participants.

These couples were first asked to have a neutral 10-minute conversation, then a 10-minute conversation about a negative topic (a characteristic of their partner that annoys them the most), followed by a 10-minute conversation about a positive topic (a characteristic of their partner that they value the most) \cite{sels2019a, sels2019b, dejonckheere2019}. 

After each conversation, each partner completed self-reports on various emotion labels such as anger, sadness, anxiety, relaxation, and happiness using a 7-point Likert scale ranging from strongly disagree (1) to strongly agree (7). Additionally, each partner watched the video recording of the conversation separately on a computer and rated his or her emotion on a moment-by-moment basis by continuously adjusting a joystick to the left (very negative) and the right (very positive), so that it closely matched their feelings, resulting in valence scores on a continuous scale from -1 to 1 \cite{gottman1985, ruef2007}. Additionally, each partner reported how they felt after the interaction and how they thought their partner felt,  using the Affect Grid questionnaire \cite{russell1989} which captures the valence and arousal dimensions of Russell’s circumplex model of emotions \cite{russell1980} resulting in values between 0 and 8 each for pleasure and arousal. 

Trained research assistants (5) listened and visually inspected the audios, and annotated the exact start and end of each talking turn for each partner. Authors that used this data categorized the valence scores into two classes, negative (0-4) and positive valence (5-8) for males and females, consequently framing the task as a binary classification task. This dataset was used by \cite{boateng2020a}

\subsection{Stanford Psychotherapy} 
Researchers collected audio, and video data from 3 heterosexual couples at Stanford University in the U.S. over 18-hour-long couple therapy sessions (18 sessions, 1 hour each for each couple A, B, and C) undergoing therapy over a period of 2 years \cite{crangle2019}. We estimated that a total of 18 hours of data was collected. No demographic information about the couples was available.

The audio was first transcribed manually, including the start and end times of each word. Then trained coders watched the video and used the transcript to code the data by marking start and end times of any of the following 4 emotions: anger, sadness, joy, tension, and neutral (defined as the absence of the 4 emotions). Each label was given a rating of low, medium, or high but the levels were not used in the analysis. These codes were adapted from Gottman’s 19 SPAFF affective codes \cite{coan2007}. To assign labels, annotators had to mark the start and end times of the occurrence of one of the emotions. Hence, no two emotion-labeled segments could overlap. Only audio data was used for recognition which was framed as a 5-class classification task. This dataset has been used by the work \cite{crangle2019}.

\subsection{UZH Couples Interactions}
Researchers collected data from 368 heterosexual German-speaking, Swiss couples (N=736 participants; age 20-80) at the University of Zurich, Switzerland over 10 years \cite{uzh2020, kuster2015}. The longitudinal study sought to investigate the impact of stress on the relationship development of couples and children across their lifespan. The average age was 47 (SD = 18.4) for women and 49 (SD = 18.2) for men, the mean relationship duration was 21 years (SD = 17.9) with 66\% being married. For women, 6\% attended the mandatory school years (9 years), 40\% completed vocational training, 21\% completed high school, and 32\% completed college or university. For men, 3\% attended the mandatory school years, 35\% completed vocational training, 12\% completed high school, and 49\% completed an academic degree.

Couples participated in three videotaped conversations in the lab each for 8 minutes — one conflict and two mutual support conversations from years 1 to 6 (one session per year). Video-recorded data from 3 couples were not available resulting in 365 couples’ data. The number of couples that took part reduced over the years with the details available at \cite{uzh2020}. Based on the data collected over the years, our estimate of the total amount of data is 637 hours.

For the conflict interaction which was used for emotion recognition, couples had to choose one problematic topic for the conflict interaction from a list of common problems (PAQ A; \cite{heavey1995}), and participants were then videotaped as they discussed the selected issue for 8 minutes. After each conversation, each partner provided self-report responses to the Multidimensional Mood questionnaire \cite{steyer1997} of their emotions on four bipolar dimensions — namely “good mood versus bad mood,” “relaxed versus angry,” “happy versus sad” and “calm versus stressed” — with the scale: 1 — very much, 2 — much, 3 — a little, 4 — a little, 5 — much, 6 — very much. The authors that used the dataset preprocessed these responses by averaging the “good mood versus bad mood” and “happy versus sad” scales and then binarized the averaged values such that values greater than or equal to 3.5 were negative (0) and the rest were positive (1).

Additionally, two research assistants were trained to code communication behaviors (interobserver agreement, k = 0.9) using an adapted version of the Specific Affect Coding System (SPAFF) \cite{coan2007}. The most prevalent code from the list was assigned every 10 seconds resulting in 48 sequences for each interaction. The codes were then grouped into positive and negative for emotion recognition.

The speech was manually annotated with the start and end of each speaker’s turn, along with pauses and noise. The speech was manually transcribed in 15-second chunks separately for each partner. Given that Swiss German is mostly spoken with different dialects across Switzerland, the spoken words were written as the corresponding German word equivalent. The following two works — \cite{boateng2021} and \cite{biggiogera2021} — used this data to recognize global and local emotions respectively, all framed as binary classification tasks.

\section{Modalities, Data Preprocessing and Feature Extraction}
In this section, we describe the modalities of the data used for emotion recognition in the surveyed works along with preprocessing approaches and features that were extracted from each modality. The surveyed papers used three distinct modalities with acoustic being the most represented modality: acoustic (19), lexical (9), and visual (2).

\subsection{Acoustic}
Given that the audio data is collected in the context of conversations, it is generally annotated manually or automatically with the segments that contain speech vs no speech (voice activity detection), and additionally, those segments are annotated to correspond to the speech of each partner, which is known as speaker diarization. 

Next, various features are extracted either using feature engineering or transfer learning. Feature engineering entails using handcrafted features that have been shown to be discriminative for the recognition task. For feature engineering, standard acoustics features such as prosodic (e.g., pitch, energy, speaking rate), spectral (e.g., mel frequency cepstral coefficients) and voice quality (shimmer and jitter) are extracted over frames of short durations (e.g., 25 ms) known as low-level descriptors (LLDs) using a sliding window (e.g., 10 ms) which may or may not be overlapping \cite{black2010, lee2010, lee2011a, lee2011b, gibson2011, katsamanis2011, black2013, lee2014, xia2015, gibson2015, li2016, li2017, tseng2018, chakravarthula2019, crangle2019, li2020, boateng2020a, boateng2021, biggiogera2021}. Various statistics called functionals (e.g., mean, median, percentiles, etc) are computed over these frames to get features for a segment (e.g., 2 seconds) or the whole audio (8-10 mins). In particular, a set of 88 features called eGeMAPS \cite{eyben2015} have been shown to be a minimalist feature set that is effective for emotion recognition tasks and have been used in the following works \cite{chakravarthula2019, boateng2021, biggiogera2021}. The openSMILE toolkit \cite{eyben2010} has been mostly used for acoustic feature extraction. Other tools such as Praat \cite{boersma2001} have been used. Due to the likelihood of having a lot of features, various features selection methods such as forward feature selection has been used \cite{black2010}. Additionally, various approaches have been used to remove the speaker, microphone, and environmental variability of the audio signal by performing mean normalizing of the LLDs for the whole session audio (e.g., \cite{black2013}). 

Transfer learning is an approach used to circumvent the need to develop hand-crafted features and entails using a model pre-trained on a different but related task (\cite{feng2020}). This process entails using the model for feature extraction or fine-tuning in which the whole model or later layers are retrained. For example, Boateng et al. took a CNN model called the YAMNET that was pretrained on an audio event classification task and used it to extract feature embeddings from spectrograms over 1-second time windows (\cite{boateng2020a}). Also, Li et al pretrained a deep learning model (CNN) on an emotion recognition task that used acted data and then used the model to extract acoustic embeddings for the recognition task \cite{li2020}.

\subsection{Lexical}
The audio is generally transcribed automatically or manually in order to use the content of the speech for emotion recognition. Various linguistic features ranging from simple features (bag of words and TF-IDF), dictionary-based features used in psychology such as LIWC \cite{pennebaker2001}, more advanced ones such as word embeddings (word2vec \cite{mikolov2013} and ELMo \cite{peters2018}) to deep learning models such as BERT \cite{devlin2018} which are currently the state-of-the-art for computing linguistic features. 

Here are examples of works that have used those lexical features: bag-of-words (unigram \cite{georgiou2011, chakravarthula2015}, ngram \cite{chakravarthula2018, chakravarthula2021, biggiogera2021}), TF-IDF \cite{katsamanis2011, lee2012, gibson2015}, LIWC \cite{biggiogera2021}, word embeddings (word2vec \cite{tseng2016, tseng2017}, ELMo \cite{chakravarthula2021}), deep sentence embeddings (seq-to-seq models \cite{tseng2017, tseng2018, tseng2019, chakravarthula2019}, BERT and Sentence-BERT \cite{boateng2021, biggiogera2021}). Transfer learning has also been used for the lexical data. For example, various sentence embeddings have been computed using pretrained models \cite{tseng2017, tseng2018, tseng2019, chakravarthula2019, boateng2021, biggiogera2021}.

\subsection{Visual}
Few works used the visual modality and in particular head movements in the videos that were recorded. The following features have been extracted: line spectral frequencies/power spectral density of head motion vectors to capture the vertical and horizontal directions of the head motion \cite{xiao2015, gibson2015}. Facial expressions were not used in those works because the quality of the video was not good enough to compute features from the face (e.g., varying sitting positions, camera distance/angle, and lighting conditions) \cite{gibson2015}. 

\section{Data Analysis and Evaluation}
In this section, we describe various algorithms that have been used for emotion recognition, multimodal fusion approaches, intrapersonal and interpersonal considerations, evaluation, and results.

\subsection{Algorithms}
The surveyed works used mostly supervised learning approaches with a few using semi-supervised \cite{lee2011b, gibson2011, katsamanis2011, lee2012, gibson2015, chakravarthula2015, xia2015} and unsupervised learning \cite{li2017, tseng2019}. The algorithms used range from simple statistical algorithms and traditional machine learning to deep learning methods. Support vector machines (SVM) have been the most used algorithm.

Here are the algorithms used by various works: SVM \cite{black2010, lee2011a, gibson2011, katsamanis2011, black2013, xia2015, xiao2015, gibson2015, li2016, tseng2016, tseng2017, boateng2020a, boateng2021, biggiogera2021}, linear discriminant analysis (LDA) \cite{black2010, xia2015}, markov models \cite{lee2010, lee2014, chakravarthula2015, xia2015, chakravarthula2018}, multiple instance learning (diversity density \cite{lee2011b, lee2012, gibson2015},  diversity density SVM \cite{gibson2011, katsamanis2011}), maximum likelihood \cite{georgiou2011, chakravarthula2015, chakravarthula2018, chakravarthula2021}, sequential probability ratio test \cite{lee2012}, logistic regression \cite{black2013}, perceptron \cite{xia2015}, gaussian mixture model (GMM) \cite{xiao2015}, deep neural networks \cite{li2016, li2017, tseng2018, chakravarthula2019}, LSTM \cite{tseng2016, tseng2017, tseng2018, tseng2019}, GRU \cite{li2020, chakravarthula2021}, random forest \cite{crangle2019, boateng2021}, CNN \cite{li2020}.

\subsection{Multimodal Fusion}
Modalities that have been combined include acoustic and lexical data \cite{katsamanis2011, tseng2018, chakravarthula2019, boateng2021}, and acoustic, lexical, and visual \cite{gibson2015}. Various fusion methods have been used such as feature-level fusion in which the features of each modality are concatenated and decision-level fusion in which each modality is trained with a separate model and the predictions from the models are combined using various approaches such as majority vote. The following papers used feature-level fusion  \cite{gibson2015, chakravarthula2019, tseng2018, boateng2021}  and decision-level fusion \cite{gibson2015, tseng2018, chakravarthula2019}. Additionally, knowledge-driven expert fusion approaches have been explored by Tseng et al. as follows: gender-based and therapy-stage fusion \cite{tseng2018}.

\subsection{Intrapersonal Considerations}
One challenge with recognizing global emotion labels — one emotion label for a long interaction duration such as 8-10 minutes — is that there is a whole range of emotions experienced and expressed throughout the interaction with different intensities. One naive approach to address this challenge is to assign every segment (e.g., 2 seconds) with the label of the whole audio and then train the model with this modified data-label pairings. This approach is used in various fields such as physical activity recognition \cite{boateng2017} since an activity label (e.g., walking) is consistent over different segments. However, such an approach is error-prone for the context of emotion recognition as it erroneously assumes that the emotion label is the same for all segments. The standard approach used in various works is to compute statistics such as mean, median, etc., over the features that have been computed over short windows as previously seen in the approach used for extracting acoustic features. 

However, the emotion recognition task may benefit from more creative modeling approaches. One such approach relates to the concept of saliency. Some interaction segments might be more salient for recognizing that one label assigned to the whole audio. Some works have leveraged some methods to identify those salient segments. One such saliency-based method is multiple instance learning which automatically identifies salient instances from a bag of instances in a semi-supervised learning fashion \cite{dietterich1997}. For example, the following works leveraged multiple instance learning to identify salient instances to use for recognition using acoustic features \cite{lee2011b, gibson2011}, lexical features \cite{lee2012}, both acoustic and lexical features \cite{katsamanis2011}, and all three modalities — acoustic, lexical, and visual \cite{gibson2015}. Another saliency-based method used the concept of the peak-end rule which posits that how people feel after an emotional experience is predicted by the emotional extremes and the end of that experience \cite{fredrickson2000}. The theory was leveraged to identify salient segments, extract features from those segments, and then perform the recognition task \cite{boateng2020a}. Another modeling approach leveraged dynamic modeling of all segments with a Markov model using acoustic features \cite{xia2015} and lexical features \cite{chakravarthula2015, chakravarthula2018}.

\subsection{Interpersonal Considerations}
The dyadic nature of couples’ interactions offers the opportunity to leverage various interaction dynamics to perform recognition of emotions. One major dyadic dynamic that has been used is synchrony/entrainment which refers to how similar/aligned/synchronized partners are when interacting. Various quantitative measures for synchrony have been computed for various modalities. For acoustic, some examples include prosodic entrainment measures computed with the following similarity measures (1) square of correlation coefficient, (2) mutual information, and (3) mean of spectral coherence over pitch and energy between the sequential turns of partner A and partner B when there is turn change \cite{lee2010}. Another approach leverages principal component analysis (PCA) to compute both prosodic and spectral entrainment while providing information about the directionality of the entrainment \cite{lee2011a, lee2014}. For the visual modality, the Kullback-Leibler (KL) divergence of the features extracted from the head motion of the partners was used as the similarity measure for synchrony \cite{xiao2015}. Another dyadic dynamic derives from the idea that partners generally influence each other while interacting. Dyadic influence has been modeled using lexical features from both partners \cite{chakravarthula2018} and both acoustic and lexical features from both partners \cite{boateng2021}. 

\subsection{Evaluation}
Three works performed regression \cite{tseng2017, tseng2018, chakravarthula2021}  with the rest performing classification. All works that performed classification trained models to perform binary classification except \cite{chakravarthula2019} and \cite{crangle2019} which performed 3-class and 5-class classification respectively. All works performed global emotion recognition except \cite{chakravarthula2019} and  \cite{biggiogera2021} which performed utterance-level recognition for every speaker turn and every 10 seconds respectively. All works have used accuracy as the evaluation metric except the following which used unweighted average recall (UAR) \cite{chakravarthula2019, boateng2020a, boateng2021, biggiogera2021}, Spearman correlation \cite{chakravarthula2021} and mean absolute error (MAE) \cite{tseng2017, tseng2018}. It is important to note that all the works that used accuracy as the metric had balanced classes (except \cite{crangle2019}) and hence, there should not be any concern about it not being an appropriate metric.

Most evaluations have been done with leave-one-couple-out (LOCO) cross-validation which is a robust evaluation approach as it gives a sense of how the model will perform on an unseen couple. With this approach, models are trained using data from all couples but one, and then the prediction is done on the remaining couple’s data as the test set. This process is repeated till each couple has been used as a test set. Hence, if there are 300 couples, the evaluation is done 300 times. In the end, the predictions of each test couple are combined either by computing the evaluation metric (e.g., accuracy) separately for each couple and then computing the mean and standard deviation of the accuracies, or concatenating all the predictions and computing one accuracy value for all the combined predictions.  LOCO is a variation of the standard leave-one-subject-out cross-validation but more robust for the context of couples data as it ensures that there is no data leakage from the same audio (as an example) being in both train and test sets. One challenge with LOCO is that the evaluation could take a long time when there are a lot of couples as it is done as many times as there are couples.

Other similarly robust evaluation approaches that have been used which also ensure that there is no data leakage but reduces the amount of time relatively are leave-N-couples-out cross-validation (LNCO) (e.g., \cite{li2020} with n=4) and K-fold cross-validation (CV) couple disjoint  (k=10: \cite{gibson2011, katsamanis2011, lee2012, boateng2021, biggiogera2021}, k=6: \cite{chakravarthula2021}). The “couple disjoint” refers to the fact that a couple is never in both the train and test set for the same evaluation run.

Another approach that has been used is the standard hold out (train test split) evaluation \cite{crangle2019}. That work also performed couple-dependent evaluation \cite{crangle2019}. That is, the authors trained and evaluated models within each couple separately. Hence, the concept of “couple disjoint” does not apply. The results from such an evaluation could be inflated as it could leverage particularities of the data to produce high results and does not give a sense of how the model will perform on an unseen couple. Nonetheless, it gives a sense of how well the model may perform if personalized models are trained.

Furthermore, several works have performed gender-specific evaluations where the model is trained and evaluated separately for male and female partners \cite{black2010, gibson2011, black2013, chakravarthula2015, xia2015, tseng2018, chakravarthula2019, crangle2019, boateng2020a, boateng2021}. The motivation for this approach is that gender differences affect how people express their emotions \cite{brody1993} and in particular how they speak and hence training approaches may benefit from using models separately for each gender. 

\subsection{Results}
In this section, we summarize the main results across the works and also provide some context to enable the correct interpretation of the results.

The best result for the work that performed 3-class classification (positive, negative, neutral) is 57.4\% UAR \cite{chakravarthula2019}. The best result for the 5-class classification (anger, sadness, joy, and tension, neutral) which also used couple-dependent evaluation ranged from 78\% to 95\% for different couples and genders \cite{crangle2019}. 

For binary classification, we provide results for 2 groups of works — (1) works that used the UCLA/UW Couples Therapy dataset in which they only considered ratings at the 2 extremes and (2) works that used other datasets without only considering the extreme ratings. We separate the two because the first task is easier than the second since only extremes are being considered rather than all ratings regardless of the intensity. 

For the first group, here are the best accuracies for each emotion task and gender with the corresponding modality shown:  
\begin{itemize}
\item Positive affect:  82\% (female) \cite{black2010} (acoustic), 78\% (male) \cite{xia2015} (acoustic), 93\% (combined male and female) \cite{katsamanis2011} (lexical)
\item Negative affect: 88.57\% (female) \cite{chakravarthula2015} (lexical), 85.7\% (male) \cite{black2013} (acoustic), 95\% \cite{katsamanis2011} (combined male and female) (lexical) 
\item Sadness: 66.4\% (female) \cite{gibson2011} (acoustic), 63.6\% (male) \cite{gibson2011} (acoustic), 80\% \cite{katsamanis2011} (combined male and female)(lexical)
\item Positive vs negative: 76\% \cite{lee2010} (combined male and female) (lexical) 
\end{itemize}

In light of these high accuracy results, it is worth noting that they are not reflective of true emotion recognition performance since the data was partitioned into two extreme ratings (top 20\% and bottom 20\%). Consequently, the performance would likely be much lower if all the data were used.

The best results for the second group are the following UAR for positive vs negative: 74.8\% (female) \cite{boateng2020a} (acoustic), 56.1\% (male) \cite{boateng2021} (acoustic and lexical), 69.2\% (combined male and female) \cite{biggiogera2021} (lexical).

For the regression tasks, the best results are 1.22 MAE for negative affect \cite{tseng2018} (acoustic and lexical) and Spearman correlation of 0.5 for positive affect, 0.58 for negative affect, 0.18 for anxiety, 0.52 for anger and 0.28 for sadness \cite{chakravarthula2021} (lexical).

For works that use gender-specific evaluations, performance for female partners tends to be better than for male partners. These results might suggest that it is more difficult recognizing the emotions of male partners and consistent with insights from psychology that suggest that female partners are more emotionally expressive \cite{brody1993}. Also, in works that consider multiple modalities, lexical modality tends to outperform other modalities including multimodal ones \cite{katsamanis2011, gibson2015, chakravarthula2019, biggiogera2021}. Considering the different evaluation contexts (e.g., different number of classes, data subsamples, etc.), it is difficult to compare results directly. Nonetheless, excluding results from UCLA/UW Couples Therapy dataset (because of the data selection bias issue) and the Stanford Lab dataset (because of the use of couple-dependent evaluation) both of which produce inflated results, it is clear that all of the best accuracy results are below 75\% with most below 70\%. As a reference, the partner-perceived result reported in Boateng et al (\cite{boateng2020a}) — how well partner A could tell the emotions of their partner B — were 73.2\% (male) and 74.3\% (female). Hence, there is more room for improvement to have performance results that are on par with or exceed how well, for example, husbands or wives could tell the emotions of their wives or husbands.

\section{Discussion: Research Gap, Challenges and Future Direction}
Despite the contributions of these works, there are still significant research gaps. In this section, we discuss these research gaps, challenges, and future directions in this area of research.

\subsection{Unexplored Modalities}
Only two modalities — acoustic and lexical — have been mostly used for recognition with the visual modality explored only superficially. Several modalities such as physiological data (heart rate, heart rate variability, skin temperature, skin conductance), body and hand gestures using accelerometer, gyroscope, or even the visual modality, and facial expression are unexplored. Additionally, only standard and simple multimodal fusion approaches have been used. More complex fusion approaches such as model-level and hybrid \cite{poria2017} could be explored in the future.

\subsection{Cross-lingual and Cross-cultural evaluation}
None of these works have performed cross-lingual and cross-cultural evaluations. Models in these works have been developed in lingual silos (English, German, and Dutch language speakers) and cultural silos (North Americans, Western Europeans). More effort would be needed to develop and evaluate recognition systems that work across languages since multilingual language models would need to be used. Furthermore, culture affects how people experience and express emotions \cite{scherer1988, matsumoto1989}. Currently, it is not clear how well the recognition systems would work across cultural contexts. Hence, more work is needed to perform these kinds of evaluations as it is important for building systems that are easily generalizable to other contexts.

\subsection{Intrapersonal and Interpersonal Modeling} 
Further intrapersonal and interpersonal modeling approaches could be explored. For example, attention mechanisms \cite{vaswani2017} could be leveraged to automatically learn the salient segments as part of the training process. Also, synchrony measures have only been computed for individual modalities. Computing and using synchrony measures multimodally is a possible future direction. Additionally, more complex dyadic influence modeling could be used such as on a turn-by-turn basis rather than only including the features of the interacting partner as was done in \cite{boateng2021}. Other kinds of dyadic dynamics from \cite{chakravarthula2020} can be used such as the ratio of both partners’ counts of positive and negative words and turn-taking patterns (e.g., the ratio of partners’ speaker turn duration, pauses, number of words, etc).

\subsection{Observed vs Self-Reported Emotion Data}
Most of these works have used observed emotion labels from external raters with only a few works using self-reported labels from the partners themselves \cite{boateng2020a, boateng2021}. One challenge with observed labels is that they are based on the perceptions of external individuals and consequently, do not reflect the subjective emotions of the partners. Though similar, performing recognition of these two groups is distinct and important to be mindful of depending on the downstream use case of the system. For example, if the intended use case and intervention is that partner A shows empathy to partner B based on how partner A is feeling, for example, a recognition system that only looks at emotional behavior or emotional expression will not be the best to use but rather, one that can adequately quantify partner A’s subjective emotions. More work is needed to be done using self-reported labels.

\subsection{Data from Daily Life}
Currently, there is no work that has performed emotion recognition using data collected from couples’ interactions in uncontrolled settings in daily life. Data from the wild tend to be noisy and could have more potential confounders such as increased heart rate arising from physical activity rather than from high emotional arousal such as stress or anger which could be the likely reason in a controlled lab conversation setting. Hence, the recognition task would be more challenging than the context of the datasets used in these works, which are couples sitting at one place, with limited mobility, and having an 8-10 minute conversation. Consequently, models developed with lab data will likely not perform well on data from daily life. Future work is needed to collect this kind of data and perform recognition with it. This work \cite{boateng2020b} is a step in that direction.

Though not unique to the context of data from daily life, it is more critical that performance evaluations go beyond the standard accuracy metrics and include detailed error analyses and assessments of conditions under which the model performs poorly. For example, the model might perform poorly when the signal-to-noise ratio is above a certain threshold (as was considered by \cite{black2010}) or if the transcript has way too few words per speaker turn. The model could be preempted from performing recognition when these conditions are encountered to reduce the likelihood of the model performing poorly. Error analyses would reveal more detailed information such as these. These are key requirements for building robust systems that work using data from uncontrolled settings.

\subsection{Real-Time Recognition Systems}
Furthermore, none of the works surveyed have actually implemented systems that perform real-time recognition either in the lab or in the real world which is the holy grail of a system for couples’ emotion recognition. The real test of such a machine learning system is its deployment and evaluation in the contexts in which they are to be used. 

Key challenges related to the turn-taking nature of couples’ interactions need to be addressed towards accomplishing this goal especially for the two most used modalities — acoustic and lexical. These include having automatic speech preprocessing pipelines — voice activity detection, speaker diarization, and speech recognition systems  — that work accurately on the fly without time lags for audio data. Several of the surveyed works used manually annotated data in the preprocessing stage and hence it is a key challenge for future works to address. 

Additionally, the recognition algorithm would also need to work accurately and without lag, on the systems that they are deployed on. For ubiquitous devices such as smartphones, smartwatches, or edge devices, the model would have to be small enough to fit on the device and perform computation without hoarding all the compute resources. Most of the works have used simple algorithms such as SVM which will work well for such contexts. But for the deep learning models, there might be potential challenges because of their size and compute requirements. This issue is more pertinent for the lexical modality since current start-of-the-art language models (e.g., BERT) are huge and might be impossible to fit on edge devices in their original form. Various approaches to compress, distill and quantize large models would need to be explored.

\section{Conclusion}
In this work, we survey 28 works that have developed and evaluated systems for emotion recognition using data collected from couples’ interactions or conversations. Overall, the works in this survey have mostly used one specific data set — UCLA/UW Couples Therapy data. All works have used data collected from lab contexts. Most works used the acoustic modality and the SVM algorithm for binary classification of positive and negative affect. Various multimodal fusion and intrapersonal and interpersonal modeling approaches have been explored. Robust evaluation approaches (e.g., LOCO, LNCO, and 10-fold CV couple disjoint) and metrics (UAR and accuracy with balanced data) have been used. Performance results leave room for improvement. Substantial research gaps remain with several opportunities for future research directions such as exploring more modalities and advanced fusion approaches, performing cross-lingual and cross-cultural evaluations, leveraging other intrapersonal and interpersonal modeling approaches, using data from daily life, and performing real-time and real-world deployment and evaluation of the recognition system. Insights from this survey would enable future research towards having better couples’ emotion recognition system that would enable social and health psychology research and the development of interventions to improve the emotional well-being, relationship quality, and chronic disease management of couples.

\begin{acks}
We are grateful to Sandeep Chakravarthula for his feedback on this paper.
\end{acks}

\bibliographystyle{ACM-Reference-Format}
\bibliography{ref}

\end{document}